\documentclass[twocolumn,superscriptaddress,
showpacs,preprintnumbers,amsmath,amssymb]{revtex4}

\usepackage{graphicx}

\newtheorem{Thm}{Theorem}
\newcommand{\bra}[1]{{\left\langle #1 \right|}}
\newcommand{\ket}[1]{{\left| #1 \right\rangle}}

\begin{document}

\title{Teleportation capability, distillability, and nonlocality on three-qubit states}

\author{Soojoon Lee}\email{level@khu.ac.kr}
\affiliation{
 Department of Mathematics and Research Institute for Basic Sciences,
 Kyung Hee University, Seoul 130-701, Korea
}
\author{Jaewoo Joo}\email{jaewoo.joo@imperial.ac.uk}
\affiliation{
 Blackett Laboratory, Imperial College London,
 Prince Consort Road, London, SW7 2BW, United Kingdom
}
\author{Jaewan Kim}\email{jaewan@kias.re.kr}
\affiliation{
 School of Computational Sciences,
 Korea Institute for Advanced Study,
 Seoul 130-722, Korea
}
\date{\today}

\begin{abstract}
In this paper, we consider
teleportation capability, distillability, and nonlocality
on three-qubit states.
In order to investigate some relations among them,
we first find the explicit formulas of the quantities about
the maximal teleportation fidelity on three-qubit states.
We show that
if any three-qubit state is useful for three-qubit teleportation
then the three-qubit state is distillable into a Greenberger-Horne-Zeilinger state,
and that
if any three-qubit state violates a specific form of Mermin inequality
then the three-qubit state is useful for three-qubit teleportation.
\end{abstract}

\maketitle

\section{Introduction}
Teleportation capability, distillability, and nonlocality have been considered as
significant features of quantum entanglement, and
have been helpful to understand quantum entanglement.
The three features have been known as a practical application of quantum entanglement,
an important method to classify quantum entanglement
with respect to the usefulness for quantum communication, and
a physical property to explain the quantum correlation, respectively.

In the case of 2-qubit states,
it has been shown that there are two relations among the three features:
If any 2-qubit state is useful for teleportation then
it is distillable into a pure entanglement, and
if any 2-qubit state violates the Bell inequality then
it is useful for teleportation~\cite{Horodeckis01}.
Then one could naturally ask what relations exist for multiqubit states.

In order to answer the question,
a proper concept of teleportation capability over multiqubit states
should be required,
since distillability and nonlocality over multiqubit systems
have been already presented,
and their relations have been appropriately investigated~\cite{Mermin,Dur,Acin,ASW01,ASW02}.
In this paper,
we present teleportation on 3-qubit states,
which could be generalized into the multiqubit case.
We then define the meaningful quantities related to the teleportation capability on 3-qubit states,
and compare the quantities with distillability and nonlocality
to look into the relations.

This paper is organized as follows.
In Sec.~\ref{Sec:2}
we properly define the quantities representing teleportation capability over 3-qubit states,
and explicitly compute the quantities.
In Sec.~\ref{Sec:3},
we show that there are two relations
among teleportation capability, distillability, and nonlocality,
which are similar to the 2-qubit case.
Finally, in Sec.~\ref{sec:4} we summarize and discuss our results.
\section{Teleportation capability over 3-qubit states}\label{Sec:2}
For teleportation over 3-qubit states,
we recall the Hillery-Bu\v{z}ek-Berthiaume~\cite{HBB} protocol,
which is the splitting and reconstruction of quantum information
over the Greenberger-Horne-Zeilinger (GHZ) state~\cite{GHZ}
by local quantum operations and classical communication (LOCC).
The protocol can be modified
into a teleportation protocol over a general 3-qubit state in the compound system $123$,
as presented in~\cite{LJK}.
\begin{figure}
\includegraphics[angle=-90,scale=0.90,width=\linewidth]{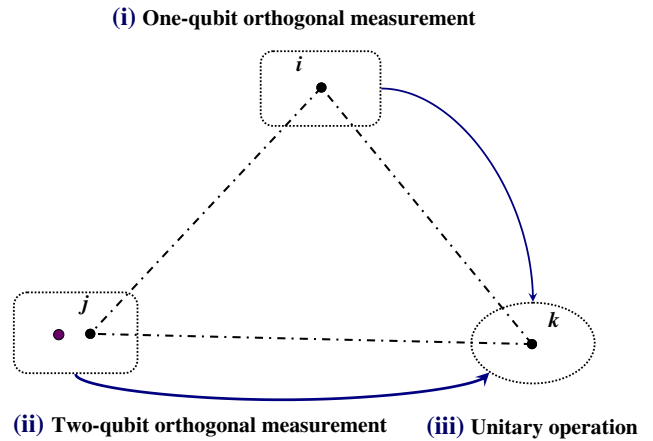}
\caption{\label{Fig:diagram}
The modified teleportation protocol over a 3-qubit state presented in~\cite{LJK}:
The dotted boxes and ellipse represent performing the orthogonal measurements
and applying the unitary operation, respectively.
The arrows represent sending classical information corresponding to the measurement results.
}
\end{figure}
The modified protocol is illustrated in Fig.~\ref{Fig:diagram}
and is described as follows:
Let $i$, $j$, and $k$ be distinct in $\{1,2,3\}$.
(i)~Make a one-qubit orthogonal measurement on the system $i$.
(ii)~Prepare an arbitrary one-qubit state,
and then make a two-qubit orthogonal measurement on the one qubit and the system $j$.
(iii)~On the system $k$,
apply a proper unitary operation
depending on the 3-bit classical information of the two above measurement outcomes.
We remark that the modified protocol is essentially equivalent to the original one
with respect to the splitting and the reconstruction of quantum information.


As mentioned in~\cite{LJK},
it is noted that any observable for a one-qubit measurement can be described as
\begin{equation}
U^{\dagger}\sigma_3U=U^{\dagger}\ket{0}\bra{0}U-U^{\dagger}\ket{1}\bra{1}U,
\label{eq:observable}
\end{equation}
where $\sigma_3=\ket{0}\bra{0}-\ket{1}\bra{1}$ is one of Pauli matrices,
and $U$ is a $2\times 2$ unitary matrix.
Thus, after the step~(i) of the teleportation protocol over a given 3-qubit state $\rho_{123}$,
the resulting 2-qubit state of the compound system $jk$ becomes
\begin{eqnarray}
\varrho_{jk}^{t}
&\equiv& \frac{\mathrm{tr}_{i}
\left(U_i^{\dagger}\ket{t}\bra{t}U_i\otimes I_{jk}
\rho_{123}
U_i^{\dagger}\ket{t}\bra{t}U_i\otimes I_{jk}\right)}
{\bra{t}U_i\rho_{i}U_i^{\dagger}\ket{t}}\nonumber\\
&=& \frac{\mathrm{tr}_{i}
\left(\ket{t}\bra{t}U_i\otimes I_{jk}
\rho_{123}
U_i^{\dagger}\ket{t}\bra{t}\otimes I_{jk}\right)}
{\bra{t}U_i\rho_{i}U_i^{\dagger}\ket{t}}
\end{eqnarray}
with probability $\bra{t}U_i\rho_{i}U_i^{\dagger}\ket{t}$
for each $t=0, 1$,
where $U_i$ is a $2\times 2$ unitary matrix of the system~$i$,
and $\rho_i=\mathrm{tr}_{jk}(\rho_{123})$.

We now review the properties of 
the teleportation fidelity~\cite{Popescu},
which represents the faithfulness of a teleportation over a 2-qubit state,
and the fully entangled fraction~\cite{BDSW,Horodeckis01,Horodeckis02,BadziagHorodeckis}.
The teleportation fidelity is naturally defined as
\begin{equation}
F(\Lambda_{\rho})=\int d\xi \bra{\xi}\Lambda_{\rho}(\ket{\xi}\bra{\xi})\ket{\xi},
\label{eq:teleportation_fidelity}
\end{equation}
where $\Lambda_\rho$ is a given teleportation protocol over a 2-qubit state $\rho$,
and the integral is performed
with respect to the uniform distribution $d\xi$ over all one-qubit pure states,
and the fully entangled fraction of $\rho$ is
defined as
\begin{equation}
f(\rho)=\max\bra{e}\rho\ket{e},
\label{eq:FEF}
\end{equation}
where the maximum is over all maximally entangled states $\ket{e}$ of 2 qubits.
It has been shown~\cite{Horodeckis02,BadziagHorodeckis} that
the maximal fidelity achievable from a given bipartite state $\rho$ is
\begin{equation}
F(\Lambda_{\rho})=\frac{2f(\rho)+1}{3},
\label{eq:2relation}
\end{equation}
where $\Lambda_{\rho}$ is the standard teleportation protocol over $\rho$ to attain the maximal fidelity.
We remark that $F(\Lambda_{\rho})>2/3$ (or $f(\rho)>1/2$)
if and only if $\rho$ is said to be useful for teleportation,
since it has been shown that the classical teleportation
can have at most $F=2/3$ (or $f=1/2$)~\cite{Horodeckis01,Popescu,MP}.

Let $F_i$ be defined as
the maximal teleportation fidelity on the resulting 2-qubit state in the compound system~$jk$
after the measurement of the system~$i$, and let $f_i$ be
the maximal average of the fully entangled fraction of the state in the compound system~$jk$
after the measurement of the system~$i$, that is,
\begin{equation}
f_i=\max_{U_i}\left[\bra{0}U_i\rho_{i}U_i^{\dagger}\ket{0}f(\varrho_{jk}^{0})
+\bra{1}U_i\rho_{i}U_i^{\dagger}\ket{1}f(\varrho_{jk}^{1})\right],
\label{eq:fi}
\end{equation}
where the maximum is over all $2\times 2$ unitary matrices.
Then, as in the 2-qubit case,
it can be obtained~\cite{LJK} that for $i\in \{1,2,3\}$
\begin{equation}
F_i=\frac{2f_i+1}{3}.
\label{eq:Fi}
\end{equation}
By the reason as in the 2-qubit case,
a given 3-qubit state $\rho_{123}$ can be said to be {\em useful for 3-qubit teleportation}
if and only if $F_i>2/3$ (or $f_i>1/2$) for every $i\in \{1,2,3\}$.

In order to explicitly calculate the values of $f_i$,
we remark that a three-qubit state $\rho_{123}$ can be described as
\begin{eqnarray}
&& \frac{1}{8}I\otimes I \otimes I\nonumber \\
 & & + \frac{1}{8}\left(\vec{s}_1\cdot \vec{\sigma} \otimes I \otimes I
+ I \otimes \vec{s}_2\cdot \vec{\sigma} \otimes I + I \otimes I \otimes \vec{s}_3\cdot \vec{\sigma}\right) \nonumber \\
&& + \frac{1}{8}\sum_{k,l=1}^3 \left(b^{kl}_1 I \otimes \sigma_k \otimes \sigma_l
+ b^{kl}_2 \sigma_k \otimes I \otimes \sigma_l + b^{kl}_3 \sigma_k \otimes \sigma_l \otimes I\right) \nonumber \\
&& + \frac{1}{8}\sum_{j,k,l=1}^3 t^{jkl} \sigma_j \otimes \sigma_k \otimes \sigma_l,
\label{eq:3qubit_state}
\end{eqnarray}
where $\sigma_i$ are Pauli matrices, $\vec{\sigma}=(\sigma_1,\sigma_2,\sigma_3)$,
$\vec{s}_i$ are real vectors in $\mathbb{R}^3$ satisfying $|\vec{s}_i|\le 1$,
and $b^{kl}_i$ and $t^{jkl}$ are real numbers.

For each $i=1,2,3$,
let $\mathbf{b}_i$ be a $3\times 3$ real matrix with $(k,l)$-entry $b^{kl}_i$.
Let $\mathbf{T}_1^j$, $\mathbf{T}_2^k$, and $\mathbf{T}_3^l$ be $3\times 3$ real matrices
with $(k,l)$-entry $t^{jkl}$, $(j,l)$-entry $t^{jkl}$, and $(j,k)$-entry $t^{jkl}$, respectively.
Then by way of the results 
in~\cite{Horodeckis01},
it is obtained that
for each $i=1,2,3$,
\begin{eqnarray}
f_i&=&\frac{1}{4}+\frac{1}{8}\max
\left[\|\mathbf{b}_i+\sum_{l=1}^3{x_l}\mathbf{T}_i^l\|+\|\mathbf{b}_i-\sum_{l=1}^3{x_l}\mathbf{T}_i^l\|\right],
\nonumber \\
\label{eq:f_i_general}
\end{eqnarray}
where $\|\cdot\|=\mathrm{tr}|\cdot|$,
and the maximum is taken over real numbers $x_l$ satisfying $x_1^2+x_2^2+x_3^2=1$.
By simple calculations of the Lagrange multiplier,
we have the following formulas:
For each $i=1,2,3$,
\begin{eqnarray}
f_i&=&\frac{1}{4}+\frac{1}{8}
\left[\|\mathbf{b}_i+\sum_{l=1}^3{y_l}\mathbf{T}_i^l\|+\|\mathbf{b}_i-\sum_{l=1}^3{y_l}\mathbf{T}_i^l\|\right],
\nonumber \\
\label{eq:f_i_general2}
\end{eqnarray}
where $y_l=\|\mathbf{T}_i^l\|/\sqrt{\sum_t\|\mathbf{T}_i^t\|^2}$.

For instance, we consider the values of $f_i$ on the class of 3-qubit states with 4 parameters
presented by D\"{u}r {\em et al.}~\cite{DCT},
\begin{eqnarray}
\rho_{\mathrm{GHZ}}&=&
\lambda_0^+\ket{\Psi_0^+}\bra{\Psi_0^+}
+\lambda_0^-\ket{\Psi_0^-}\bra{\Psi_0^-}\nonumber\\
&&+\sum_{j=1}^3 \lambda_j\left(\ket{\Psi_j^+}\bra{\Psi_j^+}+\ket{\Psi_j^-}\bra{\Psi_j^-}\right),
\label{eq:rho_GHZ}
\end{eqnarray}
where $\lambda_0^+ + \lambda_0^- + 2\sum_{j} \lambda_j =1$, and
$\ket{\Psi_j^{\pm}}=\left(\ket{j}\pm\ket{7-j}\right)/\sqrt{2}$
are the GHZ-basis states.
We note that any of 3-qubit states can be transformed
into a state $\rho_{\mathrm{GHZ}}$ in the class
by LOCC (the so-called depolarizing process)~\cite{DCT,LOCC}.

Without loss of generality,
we may assume that $\lambda_0^+$ is not less than $\lambda_0^-$ and $\lambda_j$,
since otherwise it can be adjusted by a local unitary operation.
Then by Eq.~(\ref{eq:f_i_general2}), for 4-parameter states $\rho_{\mathrm{GHZ}}$,
we obtain
\begin{eqnarray}
f_1&=&\lambda_0^+ + \lambda_3 =1/2 + (\lambda_0^+ -\lambda_0^-)/2 - \lambda_1 - \lambda_2,
\nonumber \\
f_2&=&\lambda_0^+ + \lambda_2 =1/2 + (\lambda_0^+ -\lambda_0^-)/2 - \lambda_1 - \lambda_3,
\nonumber \\
f_3&=&\lambda_0^+ + \lambda_1 =1/2 + (\lambda_0^+ -\lambda_0^-)/2 - \lambda_2 - \lambda_3.
\label{eq:f_i_ghz}
\end{eqnarray}

\section{Relations with distillability and nonlocality of 3-qubit states}\label{Sec:3}
We now take the distillability over 3-qubit states into account.
Note that if a 3-qubit state $\rho_{123}$ has
$\rho_{123}^{T_j}<0$ for all $j=1,2,3$, where $T_j$ represents the partial transposition for the system $j$,
then one can distill a GHZ state from many copies of $\rho_{123}$ by LOCC~\cite{DCT}.
(We call such a state {\em GHZ-distillable}.)
Thus, it can be obtained that a given 3-qubit state $\rho_{123}$ is GHZ-distillable
if $N_j(\rho_{123})>0$ for all $j=1,2,3$,
where
\begin{equation}
N_j(\rho_{123})=\left(\|\rho_{123}^{T_j}\|-1\right)/2,
\label{eq:N_j_def}
\end{equation}
which is called the negativity, a bipartite entanglement measure~\cite{VidalW,LCOK}.
We recall that for any 2-qubit state $\rho$,
its fully entangled fraction and its negativity satisfy the inequality
$f(\rho)\le 1/2 + N(\rho)$,
where $N$ is the negativity~\cite{VidalW}.
Then since $N$ is an entanglement monotone,
it follows from the definition of $f_i$ in Eq.~(\ref{eq:fi})
that for any 3-qubit state $\rho_{123}$
\begin{eqnarray}
f_i&=&\max_{U_i}\sum_{t=0}^1\bra{t}U_i\rho_{i}U_i^{\dagger}\ket{t}f\left(\varrho_{jk}^{t}\right)
\nonumber \\
&\le&\max_{U_i}\sum_{t=0}^1\bra{t}U_i\rho_{i}U_i^{\dagger}\ket{t}\left(1/2+N(\varrho_{jk}^{t})\right)
\nonumber \\
&\le& 1/2+N_j(\rho_{123}), \quad 1/2+N_k(\rho_{123}),
\label{eq:f_i_N_j}
\end{eqnarray}
where $i$, $j$ and $k$ are distinct in $\{1,2,3\}$.
Therefore, by the inequalities in~(\ref{eq:f_i_N_j}),
we obtain the following theorem.
\begin{Thm}\label{Thm:1}
If a 3-qubit state $\rho_{123}$ is useful for 3-qubit teleportation
then it is GHZ-distillable.
\end{Thm}
We remark that the converse of Theorem~\ref{Thm:1} is not true in general.
For example, we consider $\rho_{\mathrm{GHZ}}$
with $\lambda_0^+ = 0.4$, $\lambda_0^- =0$, and $\lambda_1=\lambda_2=\lambda_3=0.1$.
Then
since
\begin{equation}
N_j(\rho_{\mathrm{GHZ}})=\max\{0,(\lambda_0^+ -\lambda_0^-)/2 -\lambda_{4-j}\},
\label{eq:N_j}
\end{equation}
we get $N_1(\rho_{\mathrm{GHZ}})=N_2(\rho_{\mathrm{GHZ}})=N_3(\rho_{\mathrm{GHZ}})=0.1>0$,
that is, it is \textrm{GHZ}-distillable.
However, since $f_1=f_2=f_3=0.5$, it is not useful for 3-qubit teleportation.

For the nonlocality over 3-qubit states,
we consider the Mermin inequality~\cite{Mermin} on 3-qubit states.
Let $\mathcal{B}_M$ be the Mermin operator associated with the Mermin inequality as the following.
\begin{eqnarray}
\mathcal{B}_M&=&
\vec{a}_1\cdot\vec{\sigma}\otimes\vec{a}_2\cdot\vec{\sigma}\otimes\vec{a}_3\cdot\vec{\sigma}
-\vec{a}_1\cdot\vec{\sigma}\otimes\vec{b}_2\cdot\vec{\sigma}\otimes\vec{b}_3\cdot\vec{\sigma}
\nonumber\\
&&-\vec{b}_1\cdot\vec{\sigma}\otimes\vec{a}_2\cdot\vec{\sigma}\otimes\vec{b}_3\cdot\vec{\sigma}
-\vec{b}_1\cdot\vec{\sigma}\otimes\vec{b}_2\cdot\vec{\sigma}\otimes\vec{a}_3\cdot\vec{\sigma},
\nonumber\\
\label{eq:M_operator}
\end{eqnarray}
where $\vec{a}_j$ and $\vec{b}_j$ are unit vectors in $\mathbb{R}^3$.
Then for a given 3-qubit state $\rho$, the Mermin inequality is
\begin{equation}
\mathrm{tr}\left(\rho \mathcal{B}_M\right)\le 2.
\label{eq:Mermin_ineq}
\end{equation}

We take $\vec{a}_j=(0,-1,0)$ and $\vec{b}_j=(-1,0,0)$ for all $j=1,2,3$.
Then after local phase redefinition~\cite{Acin},
the Mermin operator $\mathcal{B}_M$ in Eq.~(\ref{eq:M_operator}) can be written as
\begin{equation}
\mathcal{B}_{M_0}=4\left(\ket{\Psi_0^+}\bra{\Psi_0^+}-\ket{\Psi_0^-}\bra{\Psi_0^-}\right).
\label{eq:M_operator2}
\end{equation}
Note that any 3-qubit state $\rho_{123}$ can be transformed
into a 4-parameter state $\rho_{\mathrm{GHZ}}$ in Eq.~(\ref{eq:rho_GHZ}) by the depolarizing process,
and that
$\lambda_0^{\pm}=\bra{\Psi_0^{\pm}}\rho_{\mathrm{GHZ}}\ket{\Psi_0^\pm}
=\bra{\Psi_0^{\pm}}\rho_{123}\ket{\Psi_0^\pm}$
and
$2\lambda_j=\bra{\Psi_j^{+}}\rho_{\mathrm{GHZ}}\ket{\Psi_j^+}+\bra{\Psi_j^-}\rho_{\mathrm{GHZ}}\ket{\Psi_j^-}
=\bra{\Psi_j^{+}}\rho_{123}\ket{\Psi_j^+}+\bra{\Psi_j^-}\rho_{123}\ket{\Psi_j^-}$.
Thus, for the Mermin operator $\mathcal{B}_{M_0}$ in Eq.~(\ref{eq:M_operator2}),
we obtain the following equalities:
\begin{eqnarray}
\frac{1}{4}\mathrm{tr}\left(\rho_{123} \mathcal{B}_{M_0}\right)
&=&
\bra{\Psi_0^+}\rho_{123}\ket{\Psi_0^+}-\bra{\Psi_0^-}\rho_{123}\ket{\Psi_0^-}
\nonumber\\
&=&
\bra{\Psi_0^+}\rho_{\mathrm{GHZ}}\ket{\Psi_0^+}-\bra{\Psi_0^-}\rho_{\mathrm{GHZ}}\ket{\Psi_0^-}
\nonumber\\
&=&
\lambda_0^+ - \lambda_0^-.
\label{eq:lambda_00pm}
\end{eqnarray}
We now assume that a given state $\rho_{123}$ violates the Mermin inequality
with the Mermin operator in Eq.~(\ref{eq:M_operator2}).
Then $\lambda_0^+ - \lambda_0^->1/2$, and hence
$f_i(\rho_{\mathrm{GHZ}})=\lambda_0^+ +\lambda_{4-i}>1/2$ for each $i=1,2,3$.

Since, by the definition of $f_i$ in Eq.~(\ref{eq:fi}) and Eq.~(\ref{eq:f_i_general2}),
it can be easily shown that
$f_i$ is invariant under local unitary operations
and $f_i$ is convex,
it can be also shown that
$f_i$ does not increase after applying operators in the depolarizing process~\cite{LOCC},
that is, $f_i(\rho_{\mathrm{GHZ}})\le f_i(\rho_{123})$ for each $i=1,2,3$,
and hence the given state $\rho_{123}$ is useful for 3-qubit teleportation.

Hence, if we take $\vec{a}=\vec{a}_1=\vec{a}_2=\vec{a}_3$ and $\vec{b}=\vec{b}_1=\vec{b}_2=\vec{b}_3$
then we can readily show that if the quantity
\begin{equation}
\max_{\vec{a}, \vec{b}}\mathrm{tr}\left(\rho_{123} \mathcal{B}_{M}\right)
\label{eq:max_Mermin2}
\end{equation}
is greater than 2
then $\rho_{123}$ is useful for 3-qubit teleportation.
Therefore,
we have the following relation between nonlocality and teleportation on
3-qubit states.
\begin{Thm}\label{Thm:2}
If a 3-qubit state $\rho_{123}$ violates the Mermin inequality
with respect to (\ref{eq:max_Mermin2}),
then $f_i>1/2$ for all $i=1, 2, 3$, and hence
it is useful for 3-qubit teleportation.
\end{Thm}
By Theorem~\ref{Thm:1} and Theorem~\ref{Thm:2},
we have two relations among the teleportation capability,
distillability, and nonlocality for 3-qubit states as in 2-qubit states.
\begin{figure}
\includegraphics[angle=0,scale=0.90,width=.8\linewidth]{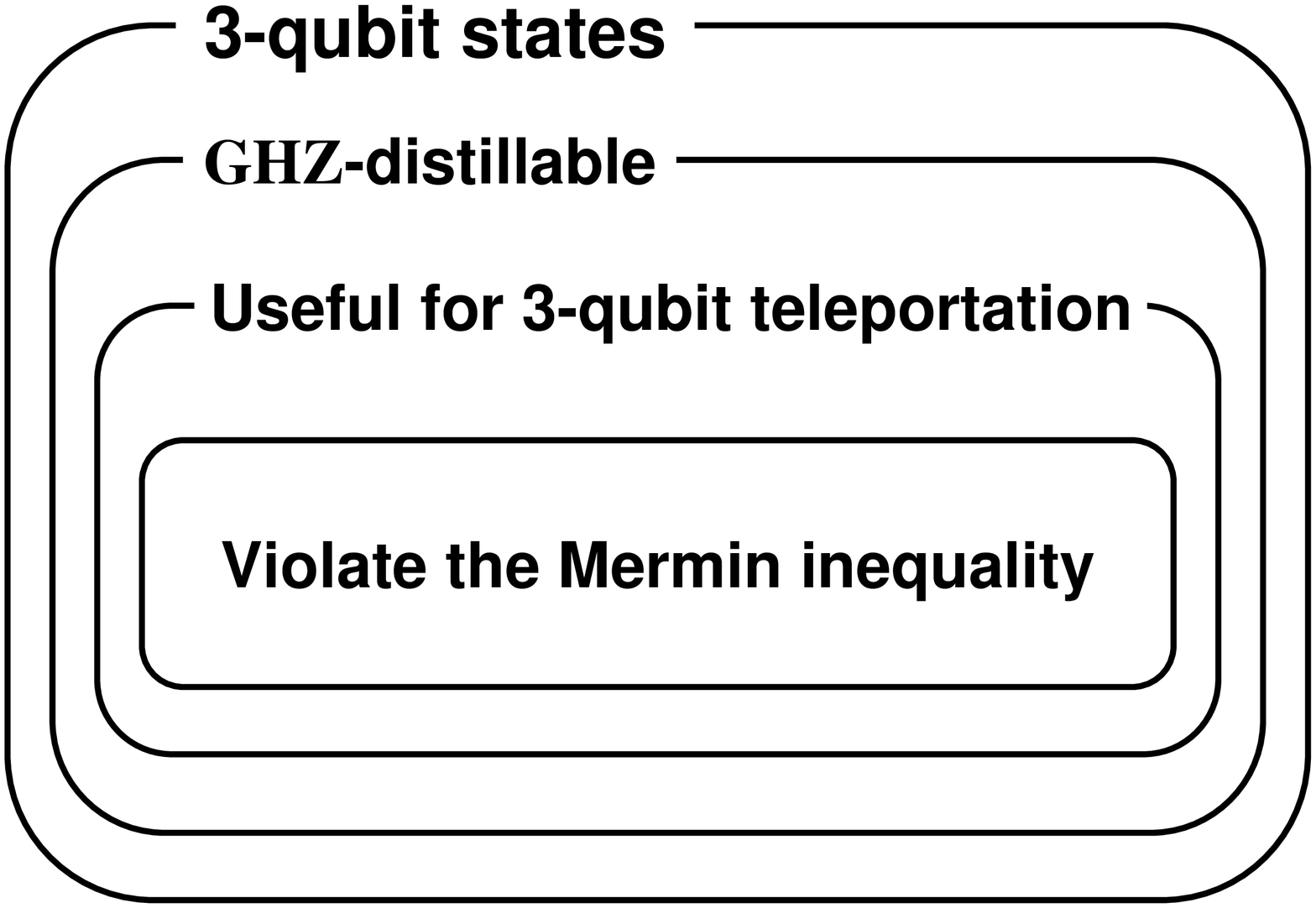}
\caption{\label{Fig:diagram2}
The relations among the teleportation capability,
distillability, and nonlocality for 3-qubit states:
The Mermin inequality we consider is
the inequality with respect to the quantity (\ref{eq:max_Mermin2}).
}
\end{figure}
The relations are seen in Fig.~\ref{Fig:diagram2}.

We remark that if we consider the Mermin inequality with respect to
the quantity
\begin{equation}
\max_{\vec{a}_j, \vec{b}_k}\mathrm{tr}\left(\rho_{123} \mathcal{B}_{M}\right),
\label{eq:max_Mermin3}
\end{equation}
where $\mathcal{B}_{M}$ is the Mermin operator in Eq.~(\ref{eq:M_operator}),
then Theorem~\ref{Thm:2} does not hold in general.
For instance, $\ket{0}(\ket{00}+\ket{11})/\sqrt{2}$ violates the Mermin inequality
with respect to (\ref{eq:max_Mermin3}),
but it is clear that the state is not useful for 3-qubit teleportation
although $f_1=1$.

However, by direct calculations
and the fact~\cite{Mermin} that the value of Eq.~(\ref{eq:max_Mermin3}) for the GHZ state is 4,
for any 4-parameter state $\rho_{\mathrm{GHZ}}$ in Eq.~(\ref{eq:rho_GHZ}),
we can explicitly find the maximum value in Eq.~(\ref{eq:max_Mermin3}),
\begin{equation}
\max_{\vec{a}_j, \vec{b}_k}\mathrm{tr}\left(\rho_{\mathrm{GHZ}} \mathcal{B}_{M}\right)
=4(\lambda_0^+ - \lambda_0^-).
\label{eq:max_Mermin4}
\end{equation}
Therefore, it can be obtained that
if $\rho_{\mathrm{GHZ}}$ violates any form of the Mermin inequality,
then it is useful for 3-qubit teleportation.

\section{Conclusions}\label{sec:4}
In conclusion,
we have considered
two relations among
the maximal teleportation fidelity,
the distillability of the GHZ state,
and violation of the Mermin inequality.
In order to investigate the relations,
we have first presented teleportation capability over 3-qubit states,
and have found the explicit formula of
the maximal teleportation fidelity
as a quantity representing the teleportation capability.
Then we have shown that
if any 3-qubit state is useful for 3-qubit teleportation
then the 3-qubit state is GHZ-distillable,
and that if any 3-qubit state violates a specific Mermin inequality
then the 3-qubit state is useful for 3-qubit teleportation.

It is known that even though for $n\ge 4$ there exist $n$-qubit bound entangled states
which violates the Mermin inequality~\cite{Dur},
there exists at least one splitting of the $n$ qubits into two groups
such that pure-state entanglement can be distilled~\cite{Acin,ASW01,ASW02}.
Therefore, if we would consider quantum communications between two or three groups
then our results could be generalized into multiqubit cases.

\acknowledgments{
S.L. was supported by the Korea Research Foundation Grant funded by the Korean Government
(MOEHRD, Basic Research Promotion Fund) (KRF-2006-331-C00044),
and J.J. by the Overseas Research Student Award Program
and the UK Engineering and Physical Sciences Research Council through the QIP IRC.
}



\begin{thebibliography}{1}
\bibitem{LOCC}
Let $\rho$ be an arbitrary 3-qubit state,
and for unitary operators $U$, $V$, $W$ in $U(2)$,
let $\mathcal{S}(U,V,W)$ be a superoperator defined as
\begin{eqnarray}
\mathcal{S}(U,V,W)(\rho)
&=&\frac{1}{2}U\otimes V \otimes W \rho U^{\dagger}\otimes V^{\dagger} \otimes W^{\dagger}
\nonumber\\
&&+ \frac{1}{2}\rho,
\label{eq:SUVW}
\end{eqnarray}
that is,
$U\otimes V \otimes W$ is performed to $\rho$ with probability 1/2,
while no operation is performed with probability 1/2.
Then we first apply $\mathcal{S}(\sigma_x,\sigma_x,\sigma_x)$ to $\rho$,
and then apply $\mathcal{S}(\sigma_z,\sigma_z,I)$,
$\mathcal{S}(\sigma_z,I, \sigma_z)$.
Then it can be easily obtained that the resulting state is diagonal in the GHZ-basis.
We now apply $\mathcal{S}(U_{3\pi/2},U_{\pi/2},I)$ and
$\mathcal{S}(U_{3\pi/2},I, U_{\pi/2})$, subsequently.
Here $U_\theta$ maps $\ket{j}$ to $e^{\iota (1-j)\theta}\ket{j}$ for $j=0,1$,
when $\iota=\sqrt{-1}$.
Then we can obtain the 3-qubit state with 4 parameters.
\end{thebibliography}
\end{document}